\documentclass[iop]{emulateapj}

\newcommand{\kms}{km\,s$^{-1}$}

\newcommand{\ci}{[\ion{C}{1}]}
\newcommand{\cii}{[\ion{C}{2}]}

\shorttitle{Herschel/HIFI observations of CO and [\ion{C}{2}] in HD 100546}
\shortauthors{Fedele et al.}

\begin{document}

\title{Probing the radial temperature structure of protoplanetary disks
  with Herschel/HIFI$^{\star}$}


\altaffiltext{$\star$}{{\it Herschel} is an ESA space observatory with science instruments provided by European-led Principal Investigator consortia and with important participation from NASA}
\author{D. Fedele\altaffilmark{1}, S. Bruderer\altaffilmark{1}, E.F. van Dishoeck\altaffilmark{1,2}, M.R. Hogerheijde\altaffilmark{2}, O. Panic\altaffilmark{3}, J. M. Brown\altaffilmark{4}, Th. Henning\altaffilmark{5} }
\altaffiltext{1}{Max Planck Institut f\"{u}r Extraterrestrische Physik, Giessenbachstrasse 1, 85748 Garching, Germany}
\altaffiltext{2}{Leiden Observatory, Leiden University, P.O. Box 9513, 2300 RA Leiden, The Netherlands}
\altaffiltext{3}{Institute of Astronomy, University of Cambridge, Madingley Road, Cambridge CB3 0HA}
\altaffiltext{4}{Harvard-Smithsonian Center for Astrophysics, 60 Garden Street, MS 78, Cambridge, MA 02138, USA}
\altaffiltext{5}{Max Planck Institute for Astronomy, K\"onigstuhl 17, 69117, Heidelberg, Germany}

\begin{abstract}
Herschel/HIFI spectroscopic observations of CO $J = 10 - 9$, CO $J = 16 - 15$ and \cii \ towards HD 100546 
are presented. The objective is to resolve the velocity profile of the lines to address the emitting region of the
transitions and directly probe the distribution of warm gas in the disk. The spectra reveal 
double-peaked CO line profiles centered on the systemic velocity, consistent with a disk origin. The
$J = 16 - 15$ line profile is broader than that of the $J = 10 - 9$ line, which in turn is broader than 
those of lower $J$ transitions ($6 - 5, 3 - 2$, observed with APEX), thus showing a clear temperature
gradient of the gas with radius. A power-law flat disk model is used to fit the CO line profiles and the CO 
rotational ladder simultaneously, yielding a temperature of $T_{\rm 0} = 1100 \pm 350\,$K (at $r_{\rm 0}$ = 13\,AU) 
and an index of $q = 0.85 \pm 0.1$ for the temperature radial gradient. 
This indicates that the gas has a steeper radial temperature gradient than the dust (mean $q_{dust} \sim 0.5$), 
providing further proof of the thermal decoupling of gas and dust at the disk heights where the CO lines form.  
The \cii \ line profile shows a strong single-peaked profile red-shifted by 0.5\,\kms \ compared to the systemic 
velocity. We conclude that the bulk of the \cii \ emission has a non-disk origin (e.g., remnant envelope or diffuse 
cloud).

\end{abstract}


\section{Introduction}

The temperature distribution of the gas in protoplanetary disks is a fundamental ingredient in models of gas and dust 
dynamics and the processes which ultimately control disk evaporation and planet formation. Thermo-chemical models 
of disks show that the gas temperature is significantly higher than that of the dust in the upper layers of the 
disk, first investigated by \citet{Kamp04} and \citet{Jonkheid04} and subsequently by many other groups 
\citep[e.g.,][]{Glassgold04, Gorti04, Nomura05, Aikawa06, Jonkheid06, Jonkheid07, Gorti08, Ercolano09, Woods09, 
Woitke09, Kamp10}. These layers also play a key role in the chemical evolution of the disk by forming molecules 
through high temperature gas-phase reactions \citep[e.g.,][]{Aikawa02,Woitke10}. However, the gas temperature in 
these layers is notoriously difficult to compute and existing models show large differences as reported by 
\citet{Roellig07} for the case of PDR models. Disk models span a  wider range in gas density and irradiation 
conditions, with even larger associated uncertainties (see discussion in \citealt{Visser12}). Addressing the 
physical properties of the gas in these layers directly through observations is thus important to test the various 
thermo-chemical disk models, which, in turn, provide the basis for the analysis of many other data. 

\noindent
Various observational studies have found evidence for vertical temperature gradients in the gas. The outer disk 
($>$ 100 AU) has been probed by single dish \citep{vanZadelhoff01} and interferometric data \citep{Dartois03, Pietu07}
of various CO low-lying pure rotational lines. For the inner disk ($<$ 50 AU), evidence for a warm gas layer with 
$T_{\rm gas} > T_{\rm dust}$ was found by \citet{Goto12} using spectrally and spatially resolved observations of CO 
ro-vibrational lines. At intermediate radii, the warm layers emit primarily at far-infrared wavelengths.
Recent observations with Herschel/PACS report the detections of pure rotational high-$J$ ($J_{\rm u}$ $>$ 12, 
$E_{\rm u}$ $\ge$ 400\,K) CO emission lines in several protoplanetary disks \citep[e.g.][]{Sturm10, vankempen10, 
Meeus12, Meeus13}. Modeling of the entire CO ladder for one source, HD 100546, clearly demonstrates the need for 
gas-dust decoupling in the upper layers throughout most of the disk \citep{Bruderer12}. However, these spatially 
unresolved PACS data do not directly probe the radial location of the warm gas. We present here spectrally resolved 
Herschel/HIFI data that uniquely determine the emitting regions through Kepler's laws.

\smallskip
\noindent
At a distance of 97\,pc \citep[$\pm$ 4\,pc;][]{vanLeeuwen07}, the Herbig Ae star HD 100546 is one of the
best studied protoplanetary disks. \citet{Bruderer12} presented a detailed analysis of the observed CO/CI/CII 
emission in this source. An unexpected conclusion was that the gaseous carbon abundance in this disk must be 
$\sim$ 5 times lower than that in the interstellar medium. Their model, however, could not reproduce the strong 
\cii \ emission detected with PACS, even after subtraction of the extended \cii \ component \citep{Sturm10,Fedele13}. 
One possibility is that the \cii \ flux measured with PACS is contaminated by a compact diffuse component, which 
can also be tested with spectrally resolved data.

\noindent
This paper presents new Herschel/HIFI observations of CO $J=10-9$, $J=16-15$ and \cii \ toward the Herbig Ae star 
HD 100546. These spectra are compared to existing APEX sub-millimeter observations of lower-$J$ CO lines  
\citep[$6-5, 3-2$, \ ][]{Panic10} with the aim of constraining the radial temperature, column density and emitting 
regions of warm CO and \cii \ in the disk and to test the predictions of thermo-chemical models. Given the spread of 
upper level energy of the four CO lines of more than one order of magnitude, with $E_{J=3}$ = 33\,K and 
$E_{J=16}$ = 751\,K, a large range of temperatures can be probed.

\begin{figure}
\epsscale{.80}
\plotone{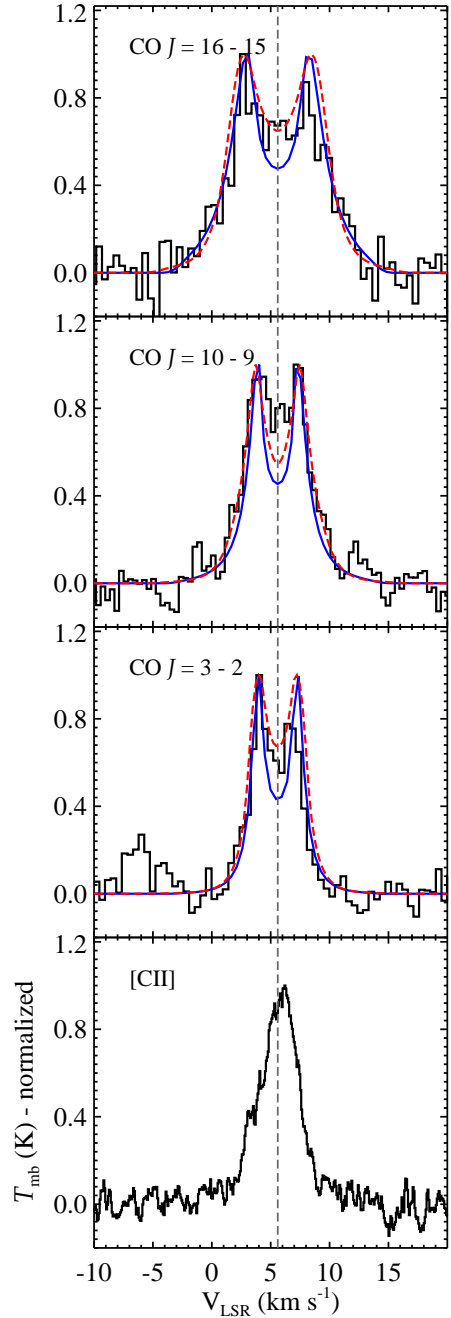}
\caption{Herschel/HIFI spectra of CO $J=16-15$, $J=10-9$ and \cii \ toward HD 100546, together with the APEX $J=3-2$ 
spectrum  (from \citealt{Panic10}). The vertical dashed line shows the system velocity. The (blue) solid line is the 
best fit power-law model (\S~\ref{sec4}) and the (red) dashed is the prediction of the thermo-chemical model 
from \citet{Bruderer12}.}
\label{fig1}
\end{figure}

\begin{deluxetable*}{llllllllllll}
\tablewidth{0pt}
\tablecaption{Parameters of the CO line emission. R$_{25\%}$ and R$_{75\%}$ indicate the radii encircling 25\% and 75\% 
of the emission as measured from the cumulative distribution (Fig.~\ref{fig3}).\label{tab:log}}
\tablehead{
\colhead{Transition}     & 
\colhead{$\nu$}          & 
\colhead{$E_{\rm u}$}      &
\colhead{HPBW}           & 
\colhead{$\Delta v_{\rm peak}$\tablenotemark{a}}  & 
\colhead{FWHM}           &
\colhead{Int. Intensity} & 
\multicolumn{3}{c}{Integrated Flux}      &
\colhead{R$_{25\%}$}      & 
\colhead{R$_{75\%}$} \\
               & [GHz]      & [K] & [\arcsec] & [\kms] & [\kms] & [K\,\kms] & \multicolumn{3}{c}{[10$^{-17}$\,W\,m$^{-2}$]} & [AU] & [AU]\\
               &            &     &           &        &        &           &  & PL\tablenotemark{c} & TC\tablenotemark{d} & & }
\startdata
CO $J$=16--15   & 1841.345   & 751 & 11.1  & 5.3 &  8.2 & 2.8 $\pm$ 0.08 &  5.9 $\pm$ 0.2 & 7.68 & 8.72 & 30 &  90 \\
CO $J$=10--9    & 1151.985   & 304 & 18.9  & 3.4 &  5.7 & 3.0 $\pm$ 0.09 &  4.5 $\pm$ 0.2 & 5.11 & 5.68 & 70 & 230 \\
CO $J$=6--5     &  691.472   & 114 &  9.1  & 2.4 &  4.2 &17.7 $\pm$ 0.9\tablenotemark{b} & 1.32 $\pm$ 0.07\tablenotemark{b} & 1.35 & 1.76 & 115 & 300 \\
CO $J$=3--2     &  345.796   &  33 & 18.2  & 2.4 &  4.0 & 4.0 $\pm$ 0.6\tablenotemark{b} & 0.15 $\pm$ 0.02\tablenotemark{b} & 0.14 & 0.26 & 150 & 320 \\
\cii \          & 1900.537   &  91 & 11.1  &     &  3.7 & 8.5 $\pm$ 0.15 & 19.6 $\pm$ 0.4 & & &     &     
\enddata
\tablenotetext{a}{Peak-to-peak velocity separation} \tablenotetext{b}{\citet{Panic10}} \tablenotetext{c}{Line flux predicted by the best-fit power-law 
model (\S~\ref{sec4}).} \tablenotetext{d}{Line flux predicted by the thermo-chemical model by \citet{Bruderer12} (\S~\ref{sec5}).}
\end{deluxetable*}

\section{Observations and data reduction}
The CO $J=16-15$ (obsid: 1342247519), CO $J=10-9$ (1342235779) and \cii \ (1342247518) observations were executed in 
dual beam switch fast chopping mode with the Wide-Band Spectrometer (WBS) and the Heterodyne Instrument for the 
Far-Infrared (HIFI) simultaneously. The spectral resolution was set to 1.1\,MHz for WBS and 0.25\,MHz for HRS for 
both polarizations. Because of the diffuse \cii \ emission, seen previously with PACS, the \cii \ observation 
of has been carried out in ``load chop'' where an internal calibration source is used in combination with an 
``off-source'' calibration observation. The half-power-beam-width (HPBW) is 18\farcs9 at the frequency of the 
CO $J=10-9$ line and 11\farcs1 at the frequencies of the CO $J=16-15$ and \cii \ lines \citep{Roelfsema12}, thus 
the beam encompasses the entire disk.

\noindent
The spectra have been extracted from the level 2 data which have been processed with standard pipeline
SPG v9.1.0. Standing waves are present in the WBS and HRS band 7 spectra of the CO $J=16-15$ and \cii \ lines. 
These have been removed by fitting a set of sine functions after masking the narrow spectral features 
(CO or \cii). This operation was performed with the 'fitHifiFringe' script provided with Hipe. 
No significant differences are observed in the two polarizations and the final spectra are obtained by averaging 
the two polarizations (for WBS and HRS separately). The data are converted from antenna temperature to mean-beam 
temperature ($T_{mb}$) dividing by $\eta_A/\eta_l$, where $\eta_A$ is the beam efficiency (0.56 for CO $J=10-9$ 
and 0.62 for CO $J=16-15$ and \cii, respectively) and $\eta_l$ is the forward efficiency, 0.96 \citep{Roelfsema12}.
The spectra are shown in Fig. \ref{fig1}.

\section{Results}
The CO $J=16-15$, CO $J=10-9$ and \cii \ lines are detected in the WBS spectrum. The \cii \ line is also detected in 
the HRS spectrum.

\subsection{CO}
The HIFI/WBS spectra of the CO lines are shown in Fig.~\ref{fig1} (top 2 panels), normalized to their peak intensity. 
The profiles of both lines show a characteristic double-peak profile created by the Keplerian motion of the gas in 
the protoplanetary disk around HD 100546. The lines are centered at V$_{\rm LSR}$ = 5.6\,\kms \ in good agreement with 
the system velocity measured using low-$J$ CO lines in the sub-millimeter \citep[e.g.][]{Panic10}. The $J=16-15$ line 
is clearly broader than the $J=10-9$ line. The integrated intensities and fluxes\footnote{To convert the integrated 
intensity ($\int{T_{mb}dV}$, K\,m\,s$^{-1}$) to integrated flux (W\,m$^{-2}$), the conversion formula is 
$2 \ k \ (\frac{\nu}{c})^3 \ \pi \Big(\frac{HPBW}{2 \ \sqrt{ln (2)}}\Big)^2 \ \int{T_{mb}dV}$, with $k$ (Boltzmann 
constant) in unit of W\,s\,K$^{-1}$, $\nu$ in Hz, c in m\,s$^{-1}$, $T_{\rm mb}$ in K.} are reported in 
Table~\ref{tab:log}. The flux of the CO $J=16-15$ line agrees well with the Herschel/PACS measurement 
($5.88 \pm 0.97 \ 10^{-17}$\,W\,m$^{-2}$, \citealt{Meeus13}). The third panel of Fig.~\ref{fig1} shows the $J=3-2$ 
sub-millimeter line, which is narrower than both higher--$J$ lines with a velocity width of only 
$\Delta v_{\rm peak} = 2.4$\,\kms \ ($\Delta v_{\rm peak}$ refers to the peak-to-peak separation).
Table~\ref{tab:log} summarizes the observed widths with the maximum velocity broadening observed in the $J=16-15$ 
line ($\Delta v = 5.3$\,\kms).  Because of high disk inclination ($42^\circ$), the Keplerian motion dominates the 
line width, meaning that the narrow lowest-$J$ line traces the slowly rotating gas in the outer part of the disk 
and the wider highest-$J$ line traces faster gas located closer to the star.

\subsection{\cii}\label{sec:cii}
The \cii \ spectrum is shown in Fig.~\ref{fig1} with a spectral resolution of 1.1\,MHz (0.17\,\kms). At this 
resolution the line does not reveal a double peaked profile characteristic of Keplerian motion. Moreover, because 
the line is centered $\sim$ 0.5\,\kms \ redward of the known system velocity, it is very likely that most of the \cii \ 
line does not come from the disk. 

\section{Analysis}\label{sec4}
The simplest possible model to analyze both the velocity profiles as well as the line intensities 
is a geometrically thin disk model in which temperature and column density are power-law functions 
of radius. The CO velocity profiles then constrain the radial temperature gradient of the gas while 
the line fluxes provide further constraints to the radial distribution of the gas column density. 
Specifically,  

\begin{equation}
N(r)  =  N_0 \ \bigg( \frac{r}{r_{\rm 0}} \bigg)^{-p}  \qquad T(r)  =  T_0 \ \bigg ( \frac{r}{r_{\rm 0}} \bigg)^{-q} 
\end{equation}

where $r_{\rm 0}$ is the inner radius of the disk, 13\,AU as found by \citet{vanderPlas09}, and $N_0$, $T_0$ 
are the column density and temperature at $r_{\rm 0}$. The motivation to use a geometrically thin disk is the 
fact that full thermochemical models \citep[e.g.][]{Bruderer12} show that the high-$J$ pure rotational CO
lines emerge from a narrow range in heights.  The free parameters of the model are $N_0, T_0, p, q$. The model 
assumptions are: 1) the gas is in Keplerian rotation; 2) CO excitation is thermalized (because of the low values 
of the line critical density, this assumption holds for all the CO lines analysed here, \citealt{Bruderer12}; 
3) the intrinsic line shape is Gaussian and the intrinsic width at each position is given by the thermal 
broadening plus turbulent broadening (fixed to $v_{\rm turb} = $0.3\,\kms). 

\subsection{Line flux from a rotating disk}
The line flux from a rotating disk is given by the integral over the disk surface

\begin{equation} \label{eq:disk_emisson}
F_\nu = \frac{\cos(i)}{d^2} \int_{r_{\rm in}}^{r_{\rm out}} \int_0^{2\pi} B_\nu(T(r)) \left(1 - e^{-\tilde{\tau}(r,\theta)}\right)  d\theta   r dr 
\end{equation}

which is defined by polar coordinates ($r$ and $\theta$). The line opacity is

\begin{equation}
\tilde{\tau}(r, \theta) = \frac{\tau(r)_{\nu - \nu/c \cdot v_{\rm proj}(r,\theta,i) } }{ \cos(i) }
\end{equation}

with $i$ ($42^\circ$) the disk inclination ($i=0^\circ$ is edge on), $d$ the distance, $B_\nu(T(r))$ the Planck 
function and $\tau(r)$ the opacity perpendicular to the disk surface. 
For a Keplerian rotating disk around a star with mass $M_*$ (2.5\,$M_{\odot}$), the projected velocity is 

\begin{equation}
v_{\rm proj}(r,\theta,i) = \sqrt{\frac{G M_*}{r}} \sin(i) \cos(\theta)  
\end{equation}

The opacity of a line connecting two CO levels $u$ and $l$ is given by
 
\begin{equation}
\tau_\nu = \frac{A_{ul} c^2}{8 \pi \nu^2} N(r) \left( x_l \frac{g_u}{g_l} - x_u \right) \times \frac{1}{\sqrt{\pi} \Delta \nu} e^{-\left(\frac{\nu}{\Delta \nu}\right)^2} 
\end{equation}

where $A_{ul}$ is the Einstein-A coefficient, $x_{l,u}$ the normalized level population, $g_{u,l}$ the 
statistical weights and $\Delta \nu$ the intrinsic line width (Doppler-parameter). 
The molecular data are from \citet{Schoier05}.
Assuming local thermodynamic equilibrium, the level population is given by $x_i=g_i \exp( - E_i / k T)/Q(T)$, with the 
level energy $E_i$ and the partition function $Q(T)$. The telescope beam is represented by a Gaussian. 
The intrinsic line width contains contributions from turbulent and thermal broadening. 
The model profiles of the three lines are shown in Fig.~\ref{fig1}.

\begin{figure}
\plotone{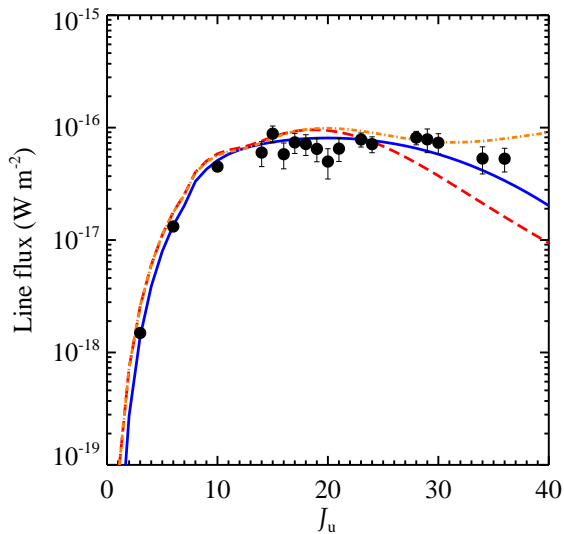}
\caption{The observed CO rotational ladder \citep{Sturm10, Meeus13} together with model fits. The (blue) solid line 
shows the best-fit power-law model, the (red) dashed line shows the prediction of the {\it standard} thermo-chemical 
model from \citet{Bruderer12} while the (orange) dot-dashed line shows the same model with a different H$_2$ formation 
rate (\S~\ref{sec5}).}\label{fig2}
\end{figure}

\begin{figure}
\centering
\plotone{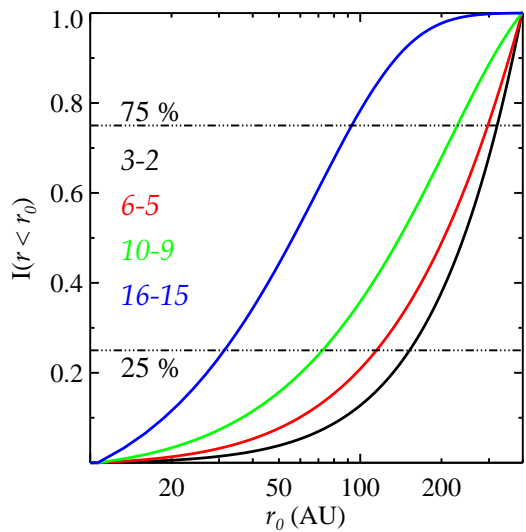}
\caption{Cumulative distribution of the 4 CO lines as predicted by the power-law model.  The dot-dashed 
lines indicate the 25\% and 75\% limits.}\label{fig3}
\end{figure}

\subsection{CO velocity profiles and rotational ladder}
The fit of the CO ladder by itself is degenerate as different power-law indices are able to reproduce the line fluxes. 
The resolved profiles of the CO lines are needed to break this degeneracy. 
The best-fit parameters are found by minimizing the $\chi^2$ (sum of the 5 individual $\chi^2$, 4 for the 
velocity profiles and 1 for the line fluxes from the CO rotational ladder). The $\chi^2$ minimization is performed 
in 2 steps: first a sparse grid of 10000 models was created for 10 different values of the parameters in the 
following ranges: $T_{\rm 0}$ between 500--1400\,K (step of 100\,K), $N_{\rm 0}$ between $10^{16}-10^{20}$\,cm$^{-2}$, 
$p,q$ between $0.5-1.4$ (step of 0.1). The minimum $\chi^2$ defines the initial guess of the best-fit parameters. 
In a second step, a denser grid of models (again 10000) is created around the initial guesses. The best-fit parameters 
are: $T_{\rm 0} = 1100 \pm 350$\,K, $q = 0.85 \pm 0.1, N_{\rm 0} = 5 \pm 3 \times 10^{17} \,{\rm cm}^{-2}, p = 0.9 \pm 0.3$.
The parameter uncertainties correspond to the 1\,$\sigma$ confidence level\footnote{The 1\,$\sigma$ confidence level 
is given by the $\Delta \chi^2 = 1$ region, which corresponds to the confidence interval of each of the four parameters 
taken separately from the others \citep[e.g.,][]{Bevington03}.} The temperature radial gradient is well constrained 
thanks to the resolved velocity profiles (values of $q < 0.6$ are excluded at 3\,$\sigma$ level). The column density 
is poorly constrained because the low-$J$ lines (up to $J = 16-15$) are optically thick \citep{Bruderer12}.

\smallskip
\noindent
The best-fit model is plotted in Figs.~\ref{fig1} and \ref{fig2} (blue solid line); the same power-law model reproduces 
the line velocity profiles and the overall shape of the rotational ladder. This simple model, however, produces too 
strong central absorption at low velocities (see discussion in \S~\ref{sec5}).

\smallskip
\noindent
Fig.~\ref{fig3} (left) shows the cumulative distribution of the different CO transitions. The dot-dashed 
lines show the 25\% and 75\% limits and the corresponding radii are given in Table~\ref{tab:log}. The values of 
$R_{25\%}$ and $R_{75\%}$ indicate the radial distances in the disk where most of the line emission comes from: 
the $J=3-2$ line emerges mainly from the outer disk ($150-320\,$AU) while the highest-$J$ line presented here 
emerges at intermediate radii ($30-90\,$AU). 

\section{Comparison to thermo-chemical models}\label{sec5}
The knowledge of the temperature and column density radial distribution of warm gas is important for our 
understanding of the disk internal structure. How do these observational results compare with gas 
temperatures of thermo-chemical models? The (red) dashed lines in Figs.~\ref{fig1} and \ref{fig2} show the 
prediction of the {\it standard} thermo-chemical model of \citet{Bruderer12} for the HD 100546 disk 
(corresponding to the representative model, discussed in \S~3 of that paper). The model results agree well 
with the observed line profiles. Specifically, it reproduces quantitatively the increasing width of the 
lines with increasing $J$. It also provides a better fit to the profiles than the power-law model at 
low-velocities because it accounts for the vertical structure of the disk. A related result of the 
full disk model is the decreasing emitting area of the lines with increasing $J$ as a consequence of 
the thermal gradient in the radial direction: \citet{Bruderer12} estimated that the 75\% of the 
$J=16-15$ line emerges from a ring between $35-80\,$AU while the $J=3-2$ line emerges from a ring between 
$70-220\,$AU, close to the inferred values.  

\smallskip
\noindent
The themo-chemical model reproduces the overall shape of rotational ladder up to the mid-$J$ lines ($J=24-23$), 
but it underestimates the flux of higher-$J$ lines. As discussed in \citet[][\S~4, 5]{Bruderer12} the flux of 
higher-$J$ lines depends on the adopted H$_2$ formation rate. In Fig.~\ref{fig2} we also show the fluxes of a 
model with a parametrized H$_2$ formation prescribed following \citet{Jonkheid04} with $T_{\rm form}=1600\,$K 
(orange dot-dashed line). This model reproduces also the higher-$J$ lines and it agrees well with the 
observed velocity profiles.

\begin{figure}
\centering
\plotone{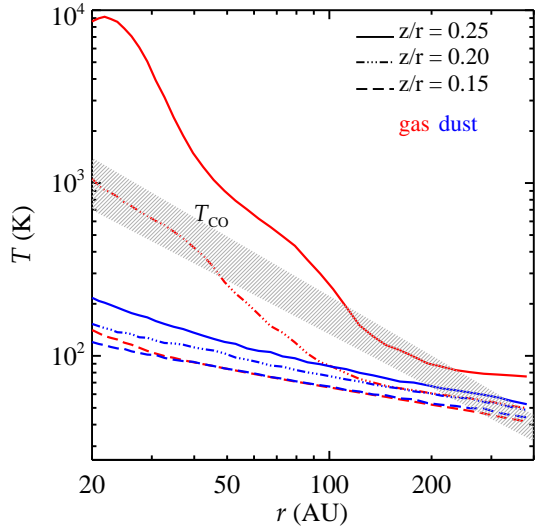}
\caption{Model predictions of the temperature radial gradient of gas (red) and dust (blue) at 3 different 
  heights in the disk of HD 100546 \citep[from][]{Bruderer12}. The dashed area shows the range 
    of best-fit power-law models.}\label{fig4}
\end{figure}

\smallskip
\noindent
The value of $q$ derived here indicates a steeper radial gradient of the gas temperature compared to the dust 
power-law index derived by interferometric observations at millimeter wavelengths \citep[mean $q_{\rm dust} \sim 0.5$, 
e.g.][]{Andrews07, Hughes08, Isella09}. The different radial gradient is likely due to the fact that the 
observed CO lines emerge from higher up in the disk compared to the mm dust emission which traces the disk mid-plane 
(where $T_{\rm gas} = T_{\rm dust}$).  In these layers, the thermo-chemical models suggest high inner temperature 
($T_0$, because of, e.g., photoelectric heating) and steep radial gradient ($q$) because of the thermal-decoupling of 
gas and dust. This is shown in Fig.~\ref{fig4} where the gas and dust temperature radial gradients are plotted for 3 
different heights ($z/r$) in the disk relevant to the CO line formation. 
The power-law model overlaps with the thermo-chemical model at $z/r=0.2$ in the inner region ($r<50\,$AU) and 
at $z/r=0.25$ in the outer disk ($r>100\,$AU).
Overall, our empirical gas temperature distribution provides a benchmark for future thermo-chemical models.

\smallskip
\noindent
Another important test for the thermo-chemical models is the nature of the \cii \ emission on HD 100546. Based 
on the non-detection of \ci, \citet{Bruderer12} speculated that the disk lacks volatile carbon (i.e., carbon 
not locked up in refractory carbonaceous grains) compared to the ISM. One option is that CO ice has been 
transformed to CH$_3$OH and more complex ice species during the cold collapse phase of the cloud. Their model 
predicts the \cii \ emission associated with the disk to be almost an order on magnitude fainter than observed 
with PACS (after subtraction of the spatially extended emission). Our finding that most of the \cii \ flux 
measured with PACS and HIFI has a non-disk origin (Fig.~\ref{fig1}, sec.~\ref{sec:cii}) is consistent with this
hypothesis.

\section{Conclusions}
In this letter, we have used spectrally resolved observations of different CO lines to constrain the radial 
temperature gradient of the warm gas in the disk around HD 100546. The spectrally resolved CO lines show a velocity 
broadening consistent with the predictions of full thermo-chemical disk models \citep{Bruderer12}. These observations 
therefore provide a new probe of the thermal decoupling of gas and dust as revealed by the high inner CO temperature 
($T_0$) and steep temperature gradient ($q$). The HIFI \cii \ spectrum confirms that the line flux measured with PACS 
is significantly contaminated by compact diffuse material.

\bibliographystyle{apj}

{\it Facilities:} \facility{Herschel}
\acknowledgments

\end{document}